# Polymerization of building blocks of life on Europa and other icy moons

Jun Kimura[1] and Norio Kitadai[1]

[1] Earth-Life Science Institute, Tokyo Institute of Technology, Tokyo 152-8550, Japan.

Corresponding author
Jun Kimura
Earth-Life Science Institute, Tokyo Institute of Technology,
2-12-1-IE-12 Ookayama, Meguro-ku, Tokyo, 152-8550, JAPAN
tel number: +81-3-5734-2851 fax number: +81-3-5734-3416
junkim@elsi.jp

**ABSTRACT**

The outer solar system may provide a potential habitat for extraterrestrial life. Remote sensing data from the Galileo spacecraft suggest that the jovian icy moons, Europa, Ganymede, and possibly Callisto, may harbor liquid water oceans underneath their icy crusts. Although compositional information required for the discussion of habitability is limited because of significantly restricted observation data, organic molecules are ubiquitous in the universe. Recently, *in-situ* spacecraft measurements and experiments suggest that amino acids can be formed abiotically on interstellar ices and comets. These amino acids could be continuously delivered by meteorite or comet impacts to icy moons. Here, we show that polymerization of organic monomers, in particular amino acids and nucleotides, could proceed spontaneously in the cold environment of icy moons, in particular the Jovian icy moon Europa as a typical example, based on thermodynamic calculations, though kinetics of formation are not addressed. Observed surface temperature on Europa is 120 and 80 K in the equatorial region and polar region, respectively. At such low temperatures, Gibbs energies of polymerization become negative, and the estimated thermal structure of the icy crust should contain a shallow region (i.e., at a depth of only a few kilometers) favorable for polymerization. Investigation of the possibility of organic monomer polymerization on icy moons could provide good constraints on the origin and early evolution of extraterrestrial life.

## 1. INTRODUCTION

The outer solar system may provide a potential habitat for extraterrestrial life. Most moons that orbit giant planets are covered with water ice and are referred to as icy moons. In addition, the detection of induced magnetic fields (Khurana et al., 1998; Kivelson et al., 2002; Khurana et al., 2008) combined with surface characteristics imaged by remote sensing for jovian icy moons by the Galileo spacecraft (Pappalardo et al., 1999) support the idea that the jovian icy moons Europa, Ganymede, and possibly Callisto may harbor liquid water oceans and possess deep habitats underneath the icy crusts. Such habitats may in some ways be similar to a terrestrial deep-sea biosphere.

However, compositional information required for the discussion of habitability is highly limited because of significantly restricted observational data. The icy crust is composed primarily of water contaminated with some non-water materials. Although some salts (e.g., hydrates of the magnesium and sodium sulfate, as well as magnesium and sodium carbonate) have been identified around tectonic features (McCord et al., 2010), further chemical environmental information remains unclear.

On the other hand, biological components are ubiquitous in the universe. Recently, glycine (Gly), the simplest amino acid, was confirmed to be present on comet 81P/Wild 2 from samples returned by NASA's Stardust spacecraft (Elsila et al., 2009). This first detection of extraterrestrial Gly suggests that amino acids can be formed by abiotic processes in the universe. In addition, production of amino acids has been experimentally confirmed in an environment of interstellar dusts and comets (Kasamatsu et al., 1997; Bernstein et al., 2002; Munos et al., 2012; Meinert et al., 2012).

During cometary impact, a non-negligible amount of certain amino acids in a comet would survive and be retained on an icy moon's surface (Pierazzo and Chyba, 2002). In fact, primitive carbonaceous meteorites with higher total concentrations of amino acids that reach 300 ppm (Burton et al., 2012) and 2,400 ppm (Pizzarello and Shock, 2010) were found. Moreover, amino acids can be produced through a process of shock synthesis from cometary material impacting the icy surface (Martins et al., 2013). If Europa's seafloor has hydrothermal systems (Vance et al., 2007), they could be candidate sites that supply amino acids to the surface through tectonic deformation of the ice crust. Furthermore, synthetic schemes for the production of amino acids in Europa's subsurface ocean have been proposed (Abbas and Schulze-Makuch, 2008). Crustal cracking could erupt salty water including amino acids from the subsurface ocean to the surface, and subsequent refreezing of the salty water could concentrate these amino acids.

In addition to amino acids, several other biochemicals have been found in extraterrestrial objects. Carbonaceous chondrites contain a variety of purines and pyrimidines, including adenine, guanine, and uracil, with total concentrations up to 500 ppb (Callahan et al., 2011; Burton et al., 2012). And long-chain monocarboxylic acids with amphiphilic properties have been extracted from the Murchison meteorite (Deamer, 1985; Sephton, 2002). Their formation could have occurred by the reaction of HCN and $NH_3$ concentrated on the parent asteroid through eutectic freezing (Miyakawa et al., 2002) or irradiation of interstellar ice with UV light (Dworkin et al., 2001). Although the detection of ribose and deoxyribose in meteorites and comets has not been reported, these compounds can be readily formed from formaldehyde through the formose reaction (Benner et al., 2012). Formaldehyde is a simple C1 compound (HCHO) that has been observed ubiquitously in interstellar clouds and comets (Ehrenfreund et al., 2002).

These organic compounds are essential building blocks of life, but subsequent evolution toward functional biopolymers remains very unclear, especially under the conditions present on icy moons. Generally, life has been characterized by the following three functions: metabolism, replication, and compartmentalization (Nakashima et al., 2001; Ruiz-Mirazo et al., 2004, 2014). In terrestrial organisms, these functions are operated by biopolymers such as protein, DNA, RNA, and phospholipids. Proteins are made of amino acids linked by peptide bonds, and DNA and RNA are made of nucleosides (composed of (deoxy) ribose and nucleobases) bound by phosphodiester linkages. Phospholipids are made of two fatty acids esterified to a glycerol phosphate molecule. These biopolymers are constructed from organic monomers by dehydration condensation reactions, as exemplified by the following equations.

2Aminoacids → Dipeptide + $H_2O$ (1)

Nucleobase + Ribose → Nucleoside + $H_2O$ (2)

2Fattyacids + Glycerolphosphate → Phospholipid + $2H_2O$ (3)

Consequently, the formations of biopolymers in liquid water is thermodynamically unfavorable. For Gly dimerization to glycylglycine (GlyGly) in water at 25 °C and neutral pH, the required Gibbs energy is 14.84 kJ $mol^{-1}$ and the equilibrium constant is $2.51\times10^{-3}$ (Kitadai, 2014), which indicates that less than 0.01% of Gly is converted into GlyGly unless the initial Gly concentration is extremely high (e.g., higher than 100 mM). The Gibbs energy decreases at higher temperatures (Shock, 1992; Kitadai, 2014). Thus, some researchers have argued that polymerization of amino acids, which is a necessary process for chemical evolution of life, occurred in submarine hydrothermal systems on primitive Earth (e.g., Imai et al. (1999); Lemke et al. (2009)).

Such reactions proceed more readily at alkaline pH (Sakata et al., 2010; Kitadai, 2014). The occurrence of high temperatures and alkaline solutions has been observed in serpentine-hosted hydrothermal systems (Kelley et al., 2001, 2005; Suda et al., 2014). Similar hydrothermal systems may prevail at the seafloor of icy moons (Vance et al., 2007). However, high temperatures also favor decomposition of organic compounds (e.g., Larralde et al. (1995); Levy and Miller (1998)). The temperature dependences of decomposition rates of amino acids are generally greater than those of polymerization (e.g., the activation energies of dimerization and decarboxylation of Gly are 88 kJ $mol^{-1}$ (Sakata et al., 2010) and 138.4 kJ $mol^{-1}$ (Li and Brill, 2003), respectively). Consequently, it is unclear whether hydrothermal systems could sustain abiotic formation of biopolymers.

In this study, we investigate polymerization of amino acids in the solid ice crust of icy moons, in particular, the Jovian icy moon Europa as a typical example, by using thermodynamic data and the revised Helgeson-Kirkham-Flowers equations of state parameters for several amino acids and their polymers updated recently on the basis of experimental thermodynamic data (Kitadai, 2014). We also calculate the Gibbs energy of formation of a nucleoside (adenosine) from adenine and ribose at low temperatures. In addition, we estimate the thermal profiles within the solid icy crust by numerically solving a heat transfer equation with observed surface temperature and the melting point of water as upper and lower boundary conditions. We evaluate the possibility of polymerization of amino acids in the surficial environment of the icy moons by examining the calculated Gibbs energies as a function of thermal structure within the crust. Although it is uncertain whether this formation of short oligomers would be enough for subsequent processes of self-organization to proceed, this type of thermodynamic analysis in icy world, which has never been previously investigated, is important for evaluating extraterrestrial abiotic synthesis.

After reviewing the environmental conditions of Europa, we describe our method for calculating the Gibbs energy of polymerization of amino acids and the thermal structure within the solid ice crust in Section 3. In Section 4, we present our main results including the calculated Gibbs energies and temperature profiles within the ice crust of Europa, and evaluate the possibilities for polymerization of amino acids in Europa's surficial environment. The possibilities for the polymerization of amino acids for other major icy moons are also discussed in Section 4. Section 5 discusses potential concentration of cometary delivered amino acids on Europa and further implications for other icy moons and asteroids. Our conclusions are summarized in Section 6.

## 2. ENVIRONMENT OF EUROPA

Europa is one of the most important candidates for hosting an extraterrestrial habitat as it may harbor a global water ocean beneath the solid icy crust. This possibility has been suggested from the detection of induced magnetic fields (Kivelson et al., 2000) combined with interpretations of imaged surface characteristics (Pappalardo et al., 1999) by remote sensing from the Galileo spacecraft and thermal equilibrium modeling (e.g., Hussmann et al., 2002). The measurement of gravity coefficients and moment of inertia factor of Europa can constrain the interior density distribution in terms of simple two- and three-layer models (Anderson et al., 1998). A two-layered interior model consists of an outer water shell and uniform silicate/metal mixed core; a three-layered model is composed of an Fe or Fe−FeS core at the center, a rock mantle surrounding the metallic core, and an outermost water shell. The thickness of the outer water shell must be approximately 80 to 170 km in both models. However, the gravity data cannot distinguish the physical state (i.e., liquid or solid) of the water shell.

Europa's atmosphere is primarily composed of molecular oxygen (Hall et al., 1995). However, it is quite tenuous, that is, 0.1 − 1 μPa (McGrath et al., 2009). Most icy moons have very thin atmospheres, too, with the exception of Sat-urn's moon Titan. Therefore, most icy moon's surfaces are exposed to sunlight and frequently impacted by small objects and energetic particles trapped within the planetary magnetic field. The surface temperature of Europa is approximately 80 K ($-193\,^{\circ}$C) and 120 K (-153 °C) in the equatorial and polar regions, respectively (Spencer et al., 1999). Europa's cold icy surface is very smooth. There are few impact craters because its surface is tectonically active and young. Based on estimates of the frequency of cometary bombardment that Europa is likely to endure, the surface is approximately $20 - 200$ million years old (Bierhaus et al., 2009). The surface is, however, covered by many tectonic features. The most striking features on Europa's surface are a series of dark lineaments and bands that crisscross the moon, which may have been produced by cracking due to tidal deformation (Greenberg et al., 1998) and icy volcanism (Figueredo et al., 2004). Another major feature of Europa is the chaos terrain, which is a locally disrupted region demonstrating a jumbled and rough texture. It has been suggested that chaos terrains are located atop lakes of liquid water within the crust (Schmidt et al., 2011). Non-water materials, particularly, hydrates of salts, are concentrated in the cracked and disrupted terrain, which have been found through the infrared mapping spectrometer onboard the Galileo spacecraft (McCord et al., 2010). This indicates that the origin of these salts is strongly related to geologic processes.

## 3. METHODS

### 3.1. Thermodynamic calculations of reactions of amino acids and nucleosides

Standard molal Gibbs energy of a given aqueous species and a crystalline compound are expressed in this study as the *apparent* standard molal Gibbs energy of formation ($\Delta G^o$), which is defined by (Helgeson et al., 1981):

$$\Delta G^o \equiv \Delta_f G^o + \left(G^o_{P,T} - G^o_{P_r,T_r}\right) \quad (4)$$

where $\Delta_f G^o$ denotes the standard molal Gibbs energy of formation of the species from the elements in their stable form at the reference pressure ($P_r$) and temperature ($T_r$) of 1 bar and 298.15 K. $G^o_{P,T} - G^o_{P_r,T_r}$ denotes the differences in the standard molal Gibbs energy of the species at pressure ($P$) and temperature ($T$) of interest, as well as those at $P_r$ and $T_r$. Thus, the Gibbs energy of formation can be calculated by using the following equations,

$$\Delta G^o = \Delta G_f^o - S_{P_r,T_r}^o(T - T_r) - c_1\left[T ln\left(\frac{T}{T_r}\right) - T + T_r\right]$$

$$- c_2\left\{\left[\left(\frac{1}{T-\Theta}\right) - \left(\frac{1}{T_r-\Theta}\right)\right]\left(\frac{\Theta - T}{\Theta}\right)\right\} + c_2\left\{\frac{T}{\Theta^2} ln\left[\frac{T_r(T-\Theta)}{T(T_r-\Theta)}\right]\right\}$$

$$+ a_1(P - P_r) + a_2 ln\left(\frac{\Psi + P}{\Psi + P_r}\right)$$

$$+ \left(\frac{1}{T-\Theta}\right)\left[a_3(P - P_r) + a_4 ln\left(\frac{\Psi + P}{\Psi + P_r}\right)\right] + \omega\left(\frac{1}{\epsilon} - 1\right)$$

$$- \omega_{P_r,T_r}\left[Y_{P_r,T_r}(T_r - T) + \frac{1}{\epsilon_{P_r,T_r}} - 1\right]$$

(5)

for aqueous species (Shock, 1992; Dick et al., 2006; Kitadai, 2014), and

$$\Delta G^o = \Delta_f G^0 - S_{P_r,T_r}^o(T - T_r) + a\left(T - T_r - T ln\left(\frac{T}{T_r}\right)\right)$$

$$- \left(\frac{(c - bT_r^2 T)(T - T_r)^2}{2T_r^2 T}\right) + V^o(\mathrm{P} - P_r)$$

(6)

for crystalline compounds (Helgeson et al., 1998; LaRowe and Helgeson, 2006; LaRowe and Dick, 2012). In Eq. 5 and 6, $S_{P_r,T_r}^o$ refers to the standard molal entropy at 298.15 K; 0.1 MPa, $a_1$, $a_2$, $a_3$, $a_4$, $c_1$, and $c_2$ denote tempera-pature- and pressure-independent parameters, and $\omega$ denotes the solvation parameters of the species of interest. $\Theta$ and $\Psi$ respectively represent solvent parameters equal to 228 K and 260 MPa (Shock, 1992). In addition, $\epsilon$ denotes the dielectric constant of $H_2O$, and $Y_{P_r,T_r}$ is the partial derivative of the reciprocal dielectric constant of $H_2O$ ($-\left(\frac{\partial(1/\epsilon)}{\partial T}\right)_P$) at 298.15 K and 0.1 MPa. The value of $Y_{P_r,T_r}$ is set to $-5.80 \times 10^{-5}$ $K^{-1}$ for all reactions (Shock et al., 1992). In Eq. 6, $V^o$ is the standard molal volume at the temperature and pressure of interest, and $a$, $b$, and $c$ are temperature-independent coefficients. In this study, values of $V^o$ are equal to those of the standard molal volumes at the reference temperature and pressure of 298.15 K and 0.1 MPa, $V^o_{P_r,T_r}$, for all crystalline compounds (LaRowe and Helgeson, 2006; LaRowe and Dick, 2012). The data and parameters required to calculate the $\Delta G^o$ as a function of temperature and pressure are presented in Table 1 and 2 for aqueous species and crystalline compounds, respectively.

The values of $\Delta G^o$ for liquid $H_2O$ (Helgeson and Kirkham, 1974) and those for $H_2O$ ice (Ih) (Feistel and Wagner, 2006) at the temperature and pressure of interest were obtained. Note that the reference state adopted in this study differs from the corresponding state used by Feistel and Wagner (2006) for $H_2O$ ice (Ih), where entropy and internal energy of liquid $H_2O$ at the triple point, T = 273.16 K and P = 611.655 Pa, were set to zero. Consequently, the $\Delta G^o$ of $H_2O$ (in both liquid and ice (Ih)) at the triple point is 0.0110 J $mol^{-1}$. Therefore, the values of $\Delta G^o$ given by Feistel and Wagner (2006) were used after adding the difference between the $\Delta G^o$ of liquid $H_2O$ at the triple point reported by Helgeson and Kirkham (1974) and that of $H_2O$ ice (Ih) reported by Feistel and Wagner (2006).

The standard state convention used for liquid $H_2O$ is one of unit activity of pure water at any temperature and pressure, whereas that for aqueous species other than $H_2O$ corresponds to the unit activity of the species in a hypothetical 1 molal solution referenced to infinite dilution at any temperature and pressure.

The standard Gibbs energies of any reaction ($\Delta_r G^o$) were calculated from the following equation:

$$\Delta_r G^o = \sum_i \nu_{i,r} \Delta G_i^o. \tag{7}$$

In equation (7), $\Delta G^o_i$ denotes the standard molal Gibbs energy of formation of the *i*-th species at any temperature and pressure and $\nu_{i,r}$ denotes the stoichiometric reaction coefficient of the *i*-th species in the reaction, which is negative for reactants and positive for products.

### 3.2. Calculation of thermal structure of ice crust

To evaluate the reactions of amino acids at the surficial regions of icy moons, we estimated the thermal structure in the ice crust. The general equation for heat transfer is given as follows:

$$\rho C_p \frac{\partial T}{\partial t} = \nabla \cdot \boldsymbol{F}, \tag{8}$$

where $\rho$ is the density, $C_p$ is the specific heat, and $\boldsymbol{F}$ is the heat flux. Equations for heat transfer, which is controlled by conduction and convection, for the solid crust are given by:

$$\boldsymbol{F} = F_{cond} + F_{conv} \tag{9}$$

$$F_{cond} = k_c \nabla T \tag{10}$$

$$F_{conv} = k_v (\nabla T - \nabla_{ad} T) \tag{11}$$

where $\nabla_{ad} T$ is the adiabatic temperature gradient and $k_c$ is the thermal conductivity, and $k_v$ is the effective thermal conductivity including the effect of thermal convection, which is given by the following.

$$k_v = \begin{cases} 0 & \frac{\partial T}{\partial r} > \left(\frac{\partial T}{\partial r}\right)_{ad} \\ \frac{\rho C_p \alpha g l^4}{18\nu}\left[\frac{\partial T}{\partial r} - \left(\frac{\partial T}{\partial r}\right)_{ad}\right] & \frac{\partial T}{\partial r} < \left(\frac{\partial T}{\partial r}\right)_{ad} \end{cases} \tag{12}$$

Here, $l$ is the characteristic length, $g$ is gravitational acceleration, $a$ is the thermal expansion coefficient, and $n$ is the local kinematic viscosity. We consider the characteristic length $l$ as the distance from the nearest boundary of the layer. The time marching scheme is the fourth-order Runge–Kutta method. This method was described in detail by Kimura et al., (2009). Surface temperature is fixed at 100 K (similar to an equatorial diurnal and seasonal mean value of 106 K (Moore et al., 2009)), and the temperature at the bottom of the crust is set to the melting point of $H_2O$ according to the crustal thickness and corresponding hydrostatic pressure. Neither tidal heating nor other heat sources are considered.

We assumed that the solid crust is composed of pure water ice and that the temperature-dependent Newtonian viscosity is as follows:

$$\eta = \eta_0 \, exp\left[A\left(\frac{T_m}{T} - 1\right)\right] \tag{13}$$

The activation parameter A is usually taken to be between 18 and 36, which corresponds to activation energy between approximately 40 and 80 kJ $mol^{-1}$. In this study, we used $A = 25$ (Goldsby and Kohlstedt, 2001). Viscosity at the melting temperature $\eta_0$ is often cited with values of approximately $10^{13}$–$10^{15}$ Pa s (e.g., Showman et al., 1997), and we used $\eta_0 = 10^{14}$ and $10^{15}$ Pa s. Other parameters for ice Ih are summarized in Table 3.

## 4. RESULTS

The standard Gibbs energies of reaction ($\Delta_r G^o$ ) for the amino acids (i.e., Gly, Alanine (Ala), and Leucine (Leu)), in particular, 2Gly → GlyGly + $H_2O$, Gly + GlyGly → GlyGlyGly + $H_2O$, Gly + Ala → AlaGly + $H_2O$, and Gly + Leu → LeuGly + $H_2O$ as a function of temperature for 1 and $10^8$ Pa corresponding to the surficial region (~1 mm depth) and at a depth of approximately 100 km, are shown in Fig.1. In a lower temperature environment, the Gibbs energies can have a negative values, that is, below −155 °C (118 K) for

2Gly → GlyGly + $H_2O$, −176 °C (97 K) for Gly + GlyGly → GlyGlyGly + $H_2O$, −193 °C (80 K) for Gly + Ala → AlaGly + $H_2O$, and below −214 °C (53 K) for Gly + Leu → LeuGly + $H_2O$, for 1 Pa. A negative Gibbs energy value means that the above reactions can proceed spontaneously.

The Gibbs energies for the above reactions at the same temperature have a positive pressure dependency (Fig.2). For the higher–pressure condition, $10^8$ Pa (Fig.1 lower panel, and Fig.2), the Gibbs energies will be slightly larger, and the temperature at which the Gibbs energy becomes negative will be lower, which means that it would be difficult for the reaction to proceed in a deeper region of the ice crust because the deeper region generally has a higher temperature. For example, the energy for polymerization of 2Gly will be negative below −171 ℃ (102 K), which means that it would be difficult for the reaction to proceed in a deeper region of the ice crust because the deeper region generally has a higher temperature (see below). Although the temperature on Europa's surface is extremely low and consequently the above reactions would be expected to proceed, the absence of atmosphere means that the surface is exposed to solar UV and local plasma in the Jovian magnetosphere, which causes decomposition. The average penetration depth for magnetospheric electrons and protons for Europa has been estimated to be 0.6 and 0.01 mm, respectively (Paranicas et al., 2009). Therefore, amino acids and their oligomers can be protected from the radiolytic decomposition even if only slightly inside the ice crust.

Figure 2 also shows the temperature profile of Europa's ice crust assuming that the surface temperature is fixed at 100 K (−173 °C) and that the thickness of the ice crust is approximately 40 km and 100 km. For a relatively thin crust, heat transfer is governed by conduction in the case of larger viscosity crust, while smaller viscosity crust is able to transfer heat by convection. A convective state thins the surficial thermal boundary layer, which means the temperature gradient will be steeper than in a conductive state. A comparison of the temperature dependence of Gibbs energies suggests that unactivated amino acid oligomerization can only occur appreciably in the shallow region, around 10 km or less, of the crust.

In parallel, the energy required for the creation of a nucleoside (adenosine) from adenine and ribose is less than that required for amino acids (Fig.3). The Gibbs energy becomes negative below −26 °C (249 K) for 1 Pa, and −131 °C (142 K) for $10^8$ Pa.

As mentioned previously, the surface temperature is a few tens of K lower at the polar region than at the equatorial region. Consequently, the 2Gly and Gly+GlyGly reactions can occur even in deeper regions of the crust. In addition, the Gly + Ala reaction can occur in the polar region, although it cannot occur on the equatorial surface. On the other hand, the higher Gibbs energy value for Gly + Leu suggests that the reaction cannot occur even in Europa's polar region. The different polymerization tendencies would bias the sequence of peptides that could be formed on the surficial region of Europa.

If Europa's icy crust is actually convecting, it is possible that polymerized amino acids and nucleosides could be transported from surficial to deep regions of the ice crust. Analysis of surface images acquired by the Galileo spacecraft suggests that there are candidate sites for subduction zones on Europa (Kattenhorn and Prockter, 2014). Several band features could have progressively removed all old cratered terrains, and surface ice and other materials could be transported into the interior. This means that surface materials can potentially be delivered to deeper regions where they would be protected from UV irradiation and all the various charged particles from Jupiter's highly radiative environment. In a deeper subsurface region, including the subcrustal water ocean, the Gibbs energies become positive, which means that the dissociation reactions (hydrolysis) will be dominant and that amino acid polymerization and nucleoside reactions are unfavorable in the deeper subsurface region. However, the kinetic inputs for hydrolysis may fall off because catalysts such as salts or cosmic ray irradiation to produce radi-cals, etc., are relatively scarce in the deeper region, which means that subsurface hydrolysis is certainly not catalytically the same as surface catalysis, and thus a mechanism for transporting surface synthesized polymers into the icy moon can be produced. In the subsurface ocean, it seems likely that, if a polymer like GlyGly absorbs onto a potentially chiral mineral or icy surface, this would decrease the solution activity of the product and drive the condensation reaction forward even under less thermodynamically favorable conditions.

The possibility of glycine polymerization on icy moons would be helpful to understand the importance of biogenic elements and prebiotic organics on Europa and whether the icy moons can have an extraterrestrial habitat or not.

## 5. DISCUSSIONS

Our calculations show that low temperature conditions, such as those that occur at the surface environment of Europa, are thermodynamically favorable for dehydration-polymerization of organic monomers. Does it follow that these reactions are kinetically possible in such an extreme environment? Unfortunately, sufficient kinetic parameters as a function of temperature are not available in the literature for these reactions; consequently, it is not possible to predict reactions rates at temperature conditions lower than 0 °C. For Gly polymerization in the solid state, many laboratory experiments have been performed with a variety of catalysts including minerals (e.g., silica (Lambert, 2008), salts (e.g., $SrCl_2$ (Kitadai et al., 2011)), and metal cations (e.g., $Cu^{2+}$ (Rode, 1999)). Interestingly, the polymerization rate of Gly does not show simple temperature dependence (i.e., slower reaction rate at lower temperature) in heterogeneous systems. For instance, Bujdak et al. (1996) conducted a heating experiment of solid Gly with montmorillonite at 70, 80, and 95 °C. After heating for 14-days, they observed that 1.13 % of the initial Gly polymerized to GlyGly at 70 °C, whereas only 0.44 and 0.59 % converted to GlyGly at 80 and 95 °C, respectively. If favorable conditions are met, polymerization of amino acids in principle can occur even at temperatures lower than 0 °C. Kaiser et al. (2013) observed formation of GlyGly and Leu-Ala at temperatures as low as 10 K under simulated interstellar ice conditions. Also, note that $MgSO_4$, a major salt identified around tectonic features of Europa's icy surface (McCord et al., 2010), has been demonstrated to accelerate Gly polymerization significantly (Kitadai et al., 2011). Based on these reported experimental results, it is possible that the kinetic and thermodynamic construction of biopolymers is feasible in Europa's surface environment.

Moreover, exogenously delivered amino acids should be considered. First, what quantity of amino acids could be delivered exogenically onto the moon? It has been estimated that 90% or more of the craters on the Galilean satellites are due to impact by Jupiter-family comets, with Trojan asteroids and long period comets being the second and third major impactor types (Zahnle et al., 1998). The average impact velocity of these comets on Europa is calculated to be around 26 km $s^{-1}$, with 10% of the objects striking Europa at velocities below 16 km $s^{-1}$ (Zahnle et al., 1998). In such impact incidents on Europa, the surviving fractions of amino acids have been estimated to be 0–5.26% and 0.45–16.94% for the average and lower impact velocities, respectively. Although the abundance of amino acids in comets is not known, it has been suggested that comets contain about 10 times the amount of organics in carbonaceous chondrites (Delsemme 1991; Alexander et al., 2007). Carbonaceous chondrites show considerable concentration variations of amino acids ranging from 0.2 ppm (CB chondrite; Burton et al., 2013) to 2,400 ppm (CR2 chondrite; Pizzarello and Shock, 2010). CR2s are the most primitive, least altered chondrites (Cody and Alexander, 2005; Alexander et al., 2010), and generally have higher concentration of amino acids than chondrites that experienced more severe aqueous or thermal alteration in meteorite parent bodies (Burton et al., 2012). Organic matter in carbonaceous chondrites and comets has been inferred to be formed from a common precursor material that formed in the outer solar system and/or in the interstellar medium (Alexander et al., 2007; Herd et al., 2011). Therefore, amino acid concentration in comets could be as high as 24,000 ppm. Based on these estimates, together with the current impactor mass flux at Europa of $5.5 \times 10^4$ kg yr$^{-1}$ (Pierazzo and Chyba, 2002), the amount of amino acids delivered on the Europa's surface is calculated to be the order of $1 \times 10^2$ kg yr$^{-1}$. If the input of amino acids continues for 50 Myr (a typical surface age based on the size distribution analysis of impact craters by Bierhaus et al., (2009)), the overall delivery of cometary amino acids is $5 \times 10^9$ kg. If this input is distributed evenly on Europa's surface, and is mixed homogenously with a depth of 1 meter of the ice shell (an average gardening depth of 10 Myr-old (Carlson et al., (2009)), the resultant surface concentration of amino acids is around 0.1 ppm. In contrast, localized impact events associated with re-freezing of surface ice could afford higher local concentration of amino acids. For instance, a 0.9-km diameter comet of density 800 kg m$^{-3}$ ($3.1 \times 10^{11}$ kg) forms a 20 km-size crater (Pierazzo and Chiba, 2002). Concentration of delivered amino acids could reach the order of 1,000 ppm if the amino acids are retained within the crater region. Future in-situ measurements with advanced analytical setups are expected to provide an exact amount of amino acids on Europa's surface.

In addition, Europa's thermodynamically favorable environment for dehydration–condensation of organic monomers is also likely for other icy moons. The surface temperature of moons without an atmosphere generally decreases with increasing distance from the Sun. On Enceladus, a Saturnian icy moon, the observed surface temperature is 32.9 K at the north pole and 145 K at the south pole, where water vapor, icy particles, and organic compounds are sprayed into space, and 75 K as a disk-integrated value (Spencer et al., 2006). Triton, the largest Neptunian moon, has a surface temperature of 38 K (McKinnon et al., 2007). These colder environments are more favorable for dehydration–condensation than Europa's surface; thus, condensation reactions can occur in the deeper region of the ice crust. Furthermore, the condensation of Gly+Leu, which can not take place thermodynamically in Europa's environment, can proceed in Triton's.

Titan is a unique icy moon with a thick nitrogen–dominated atmosphere. The surface of Titan is covered primarily by water ice, and the ground temperature and pressure have been measured as 93.7 K and 1.47 bars, respectively, by the Huygens probe (Fulchignoni et al., 2005). General circulation model results suggest that seasonal and latitudinal variation of air temperatures near the surface is small. This variation is estimated to be less than 1 K at low latitudes and up to 4 K at the poles, with a maximum temperature of 94 K during the southern summer solstice. In addition, a minimum temperature of 90 K is estimated during the polar winter (Tokano and Neubauer, 2002). Therefore, according to our results, the reactions of 2Gly and Gly+GlyGly can take place, similar to Europa. In addition, nucleotide bases and amino acids have been found experimentally in a simulated Titan atmosphere (Hörst et al., 2012). Titan's surface temperature and pressure are close to the triple point of methane (90.7 K) and ethane (90.4 K); thus, these materials can co-exist in solid, liquid, and vapor phases. The presence of liquid lakes of hydro-carbons on the surface has been confirmed (Stofan et al., 2007), and some lakes can possibly dry up by subsurface hydraulic flow or evaporation (Hayes et al., 2008). In addition, new lakes that have appeared after a large outburst of clouds have also been detected (Turtle et al., 2009), which suggests that some of the lakes are transient and appear to be related to atmospheric global circulation of methane, which is considered analogous to Earth's water cycle, although at a much lower temperature. Such repetitive flooding and drying processes could have provided a driving force for increments in concentration and cyclic replication of prebiotic precursor molecules, which is similar to Earth's early tidal cycle (Lathe, 2004).

It should be noted that the reactions discussed above are not limited to an icy environment and can also take place on a rocky surface if the temperature is sufficiently low so that the Gibbs energy becomes negative. For example, organic material in primitive chondritic meteorites is generally highly dehydrated despite the fact that the petrology of meteorites retain a record of aqueous alteration in the meteorite parent bodies (e.g., Pizzarello et al. (2006)). The origin, mechanism of alteration, and timing of the organic matter has been a long-standing problem in astrochemistry, whereas studies for these topics are extensive (e.g., Quirico et al. (2014)). It has been estimated that the main belt asteroid Vesta has an annual mean surface temperature range of 176–188 K (Titus et al., 2012). The interior temperature of small asteroids (and meteorites) is expected to be much lower and possibly fits into the polymerization-allowing temperature window as shown in this study. Our calculation therefore opens the possibility that organic matter in meteorites has been subjected to continuous dehydration after parent body processings over 4.5 billion years.

## 6. SUMMARY

In this study, we calculated the energetics of polymerization of amino acids and the production of a nucleoside (adenosine) in the environments of Europa and other icy moons. Our major finding are as follows:

• In a low temperature environment, the Gibbs energies of these reactions can become negative. For example, for 2Gly→GlyGly+$H_2O$, reactions can proceed spontaneously at temperatures below 118 K (−155 °C). The pressure dependencies of the reactions of amino acids are relatively small.

• The surface temperature on Europa is approximately 80 K (−193 °C) around the pole and approximately 120 K (−153 °C) at the equatorial regions. From the temperature profiles in the ice crust, it is expected that the Gibbs energy will be negative only in the shallow region (i.e., at a depth of only a few kilometers), whereas the temperature in deeper regions is not favorable for polymerization.

• Europa and other icy moons have colder surfaces and therefore have the potential to facilitate such reactions.

## ACKNOWLEDGEMENTS

The authors would like to thank anonymous reviewers for their detailed comments and suggestions. Insightful comments from Henderson J. Cleaves significantly improved this paper. This work of J. K. is supported by the Grants-in-Aid for Scientific Research (KAKENHI) from the Ministry of Education, Culture, Sports, Science and Technology (25800242).

## REFERENCES

Abbas, S. H., and Schulze-Makuch, D. (2008) Amino acid synthesis in Europa's subsurface environment. *Int. J. Astrobiology* 7:193–203.

Alexander, C. M. O. D., Fogel, M., Yabuta, H., and Cody, G. D. (2007) The origin and evolution of chondrites recorded in the elemental and isotopic compositions of their macromolecular organic matter. *Geochim. Cosmochim. Acta* 71:4380–4403.

Alexander, C. M. O 'D., Newsome, S. D., Fogel, M. L., Nittler, L. R., Busemann, H., and Cody, G. D. (2010) Deuterium enrichments in chondritic macromolecular material – implications for the origin and evolution of organics, water and asteroids. *Geochim. Cosmochim. Acta* 74:4417–4437.

Anderson, J. D., Schubert, G., Jacobson, R. A., Lau, E. L., Moore, W. B., and Sjogren W. L. (1998) Europa's differentiated internal structure: Inferences from four Galileo encounters. *Science* 281:2019–2022.

Benner, S. A., Kim, H.-J., and Carrigan, M. A. (2012) Asphalt, water, and the prebiotic synthesis of ribose, ribonucleosides, and RNA. *Acc. Chem. Res.* 45:2025–2034.

Bernstein, M. P., Dworkin, J. P., Sandford, S. A., Cooper, G. W., and Allamandola, L. J. (2002) Racemic amino acids from the ultraviolet photolysis of interstellar ice analogues, *Nature* 416:401–403.

Bierhaus, E. B., Zahnle, K., and Chapman, C. R. (2009) Europa's crater distributions and surface ages. In *Europa*, edited by R. T. Pappalardo, W. B. McKinnon and K. Khurana, The University of Arizona Press, Tucson, AZ, pp 161–180.

Bujdak, J., Son, H. L., Yongyai, Y., and Rode, B. M. (1996) The effect of reaction conditions on montmorillonite-catalyzed peptide formation. *Catalysis Lett.* 37:267–272.

Burton, A. S., Stern, J. C., Elsila, J. E., Glavin, D. P., and Dworkin, J. P. (2012) Understanding prebiotic chemistry through the analysis of extraterrestrial amino acids and nucleobases in meteorites. *Chem. Soc. Rev.* 41:5459–5472.

Burton, A. S., Elsila, J. E., Hein, J. E., Glavin, D. P., and Dworkin, J. P. (2013) Extraterrestrial amino acids identified in metal-rich CH and CB carbonaceous chondrites from Antarctica. *Meteor. Planet. Sci.* 48:390–402.

Callahan, M. P., Smith, K. E., Cleaves, H. J., Ruzicka, J., Stern, J. C., Glavin, D. P., House, C. H., and Dworkin, J. P. (2011) Carbonaceous meteorites contain a wide range of extraterrestrial nucleobases. *Proc. Natl. Acad. Sci. USA* 108:13995–13998.

Carlson, W. M., Calvin, W. M., Dalton, J. B., Hansen, G. B., Hudson, R. L., Johnson, R. E., McCord, T. B., and Moore, M. H. (2009) Europa's surface composition. In *Europa*, edited by R. T. Pappalardo, W. B. McKinnon and K. Khurana, The University of Arizona Press, Tucson, AZ, pp. 283–328.

Cody, G. D., and Alexander, C. M. O 'D. (2005) NMR studies of chemical structural variation of insoluble organic matter from different carbonaceous chondrite groups. *Geochim. Cosmochim. Acta* 69:1085–1097.

Cox, J. D., Wagman, D. D., and Medvedev, V. A. (1989) *CODATA Key Values for Thermodynamics*, Hemisphere Publishing Corporation, New York.

Deamer, D. W. (1985) Boundary structures are formed by organic components of the Murchison carbonaceous chondrite. *Nature* 317:792–794.

Diaz, E. L., Domalski, E. S., and Colbert, J. C. (1992) Enthalpies of combustion of glycylglycine and DL-alanyl-DL-alanine. *J. Chem. Thermodyn.* 24:1311–1318.

Dick, J. M., LaRowe, D. E., and Helgeson, H. C. (2006) Temperature, pressure, and electrochemical constraints on protein speciation: Group additivity calculation of the standard molal thermodynamic properties of ionized unfolded proteins. *Biogeosci.* 3:311–336.

Delsemme, A. H. (1991) Nature and history of the organic compounds in comets: An astrophysical review. In *Comets in the Post-Halley Era*, edited by

R. L. Newburn, M. Neugenbauer and J. Rahe, Vol. 1, Kluwer Academic, Dordrecht, pp 377–428..

Dworkin, J. P., Deamer, D. W., Sandford, S. A., and Allamandola, L. J. (2001) Self-assembling amphiphilic molecules: Synthesis in simulated interstellar/precometary ices. *Proc. Natl. Acad. Sci. USA* 98:815–819.

Ehrenfreund, P., Irvine, W., Becker, L., Blank, J., Brucato, J. R., Colangeli, L., Derenne S., Despois, D., Dutrey, A., Fraaije, H., Lazcano, A., Owen, T., Robert, F., and an International Space Science Institute ISSI-Team. (2002) Astrophysical and astrochemical insights into the origin of life. *Rep. Prog. Phys.* 65:1427–1487.

Elsila, J. E., Glavin, D. P., and Dworkin, J. P. (2009) Cometary glycine detected in samples returned by Stardust, *Meteor. Pla. Sci.* 44:1323–1330.

Feistel, R., and Wagner, W. (2006) A new equation of state for $H_2O$ ice Ih. *J. Phys. Chem. Ref. Data* 35:1021–1047.

Figueredo P. H., and Greeley, R. (2004) Resurfacing history of Europa from pole-to-pole geological mapping. *Icarus* 167:287–312.

Fulchignoni, M., Ferri, F., Angrilli, F., Ball, A. J., Bar-Nun, A., Barucci, M. A., Bettanini, C., Bianchini, G., Borucki, W., Colombatti, G., Coradini, M., Coustenis, A., Debei, S., Falkner, P., Fanti, G., Flamini, E., Gaborit, V., Grard, R., Hamelin, M., Harri, A. M., Hathi, B., Jernej, I., Leese, M. R., Lehto, A., Lion Stoppato, P. F.,

Lopez-Moreno, J. J., Makinen, T., McDonnell, J. A. M., McKay, C. P., Molina-Cuberos, G., Neubauer, F. M., Pirronello, V., Rodrigo, R., Saggin, B., Schwingenschuh, K., Seiff, A., Simoes, F., Svedhem, H., Tokano, T., Towner, M. C., Trautner, R., Withers, P., and Zarnecki, J. C. (2005) In situ measurements of the physical characteristics of Titan's environment. *Nature* 438,:785-791.

Goldsby, D. L., and Kohlstedt, D. L. (2001) Superplastic deformation of ice: Experimental observations. *J. Geophys. Res.* 106:11017–11030.

Greenberg, R., Geissler, P., Hoppa, G., Tufts, B. R., Durda, D. D., Pappalardo, R. T., Head, J. W., Greeley, R., Sullivan, R., and Carr, M. H. (1998) Tectonic processes on Europa: Tidal stresses, mechanical response, and visible features. *Icarus* 135:64–78.

Hall, D. T., Strobel, D. F., Feldman, P. D., McGrath, M. A., and Weaver, H. A. (1995) Detection of an oxygen atmosphere on Jupiter's moon Europa. *Nature* 373:677–681.

Hayes, A., Aharonson, O., Callahan, P., Elachi, C., Gim Y., Kirk R., Lewis, K., Lopes, R., Lorenz, R., Lunine, J., Mitchell, K., Mitri, G., Stofan, E., and Wall S. (2008) Hydrocarbon lakes on Titan: Distribution and interaction with a porous regolith. *Geophys. Res. Lett.* 35:L9204.

Helgeson, H. C., and Kirkham, D. H. (1974) Thermodynamic prediction of the thermodynamic behavior of aqueous electrolytes at high pressures and temperatures: I. Summary of the thermodynamic/electrostatic properties of the solvent. *Am. J. Sci.* 274:1089–1198.

Helgeson, H. C., Kirkham, D. H., and Flowers, G. C. (1981) Theoretical prediction of the thermodynamic behavior of aqueous electrolytes at high pressures and temperatures: IV. Calculation of activity coefficients, osmotic coefficients, and apparent molal and standard and relative partial molal properties to 600° C and 5 kb. *Am. J. Sci.* 281:1249–1516.

Helgeson, H. C., Owens, C. E., Knox, A. M., and Richard, L. (1998) Calculation of the standard molal thermodynamic properties of crystalline, liquid, and gas organic molecules at high temperature and pressures. *Geochim. Cosmochim. Acta* 62:985–1081. Hobbs, P. V. (1974) Ice Physics. Oxford University Press, London, UK.

Herd, C. D. K., Blinova, A., Simkus, D. N., Huang, Y., Tarozo, R., Alexander, C. M. O'D., Gyngard, F., Nittler, L. R., Cody, G. D., Fogel, M. L., Kebukawa, Y., Kilcoyne, A. L. D., Hilts, R. W., Slater, G. F., Glavin, D. P., Dworkin, J. P., Callahan, M. P., Elsila, J. E., De Gregorio, B. T., and Stroud, R. M. (2011) Origin and evolution of prebiotic organic matter as inferred from the Tagish lake meteorite. Science 332:1304–1307.

Hörst, S. M., Yelle, R. V., Buch, A., Carrasco, N., Cernogora, G., Dutuit, O., Quirico, E., Sciamma-O'Brien, E., Smith, M. A., Somogyi, A., Szopa, C., Thissen, R. and Vuitton, V. (2012) Formation of amino acids and nucleotide bases in a Titan atmosphere simulation experiment. *Astrobiology* 12:1–9.

Hussmann, H., Spohn, T., and Wieczerkowski, K. (2002) Thermal equilibrium states of Europa's ice shell: Implications for internal ocean thickness and surface heat flow. *Icarus* 156:43–151.

Imai, E., Honda, H., Hatori, K., Brack, A., and Matsuno, K. (1999) Elongation of oligopeptides in a simulated submarine hydrothermal system. *Science* 283:831–833.

Kaiser, R. I., Stockton, A. M., Kim, Y. S., Jensen, E. C., and Mathies, R. A. (2013) On the formation of dipeptides in interstellar model ices. *Astrophys. J.* 765:111.

Kasamatsu, T., Kaneko, T., Saito, T., and Kobayashi, K. (1997) Formation of organic compounds in simulated interstellar media with high energy particles. *Bull. Chem. Soc. Jpn.* 70:1021–1026.

Kattenhorn, S. A., and Prockter, L. M. (2014) Evidence for subduction in the ice shell of Europa. *Nature Geosci.* 7:762-767.

Kelley, D. S., Karson, J. A., Blackman, D. K., Früh-Green, G. L., Butterfield, D. A., Lilley, M. D., Olson, E. J., Schrenk, M. O., Roe, K. K., Lebon, G. T., Rivizzigno, P., and AT3-60 Shipboard Party. (2001) An off-axis hydrothermal vent field near the Mid-Atlantic Ridge at 30 degrees N. *Nature* 412:145–149.

Kelley, D. S., Karson, J. A., Früh-Green, G. L., Yoerger, D. R., Shank, T. M., Butterfield, D. A., Hayes, J. M., Schrenk, M. O., Olson, E. J., Proskurowski, G., Jakuba, M., Bradley, A., Larson, B., Ludwig, K., Glickson, D., Buckmen, K., Bradley, A. S., Brazelton, W. J., Roe, K., Elend, M. J., Delacour, A., Bernasconi, S. M., Lilley, M. D., Baross, J. A., Summons, R. E., and Sylva, S. P. (2005) A serpentine-hosted ecosystem: The lost city hydrothermal field. *Science* 307:1428–1434.

Khurana, K. K., Kivelson, M. G., Stevenson, D. J., Schubert G., Russell, C. T., Walker R. J., and Polanskey, C. (1998) Induced-magnetic fields as evidence for subsurface oceans in Europa and Callisto. *Nature* 395:777–780.

Khurana, K. K., Kivelson, M. G., Hand, K. P., and Russel, C. T. (2008) Electromagnetic induction from Europa's ocean and the deep interior. In *Europa*, edited by R. T. Pappalardo, W. B. McKinnon and K. Khurana, The University of Arizona Press, Tucson, AZ, pp. 571–586.

Kimura, J., Nakagawa, T., and Kurita K. (2009) Size and compositional constraints of Ganymede's metallic core for driving an active dynamo. *Icarus* 202:216–224.

Kitadai, N., Yokoyama, T., and Nakashima, S. (2011) Hydration-dehydration interactions between glycine and anhydrous salts: Implications for a chemical evolution of life. *Geochim. Cosmochim. Acta* 75:6285-6299.

Kitadai, N. (2014) Thermodynamic prediction of glycine polymerization as a function of temperature and pH consistent with experimentally obtained results. *J. Molecular Evolution* 78:171–187.

Kivelson, M. G., Khurana, K. K., Russell, C. T., Volwerk, M., Walker, R. J., and Zimmer, C. (2000) Galileo magnetometer measurements: A stronger case for a subsurface ocean at Europa. *Science* 289:1340–1343.

Kivelson, M. G., Khurana, K. K., and Volwerk, M. (2002) The permanent and inductive magnetic moments of Ganymede. *Icarus* 157:507–522.

LaRowe, D. E., and Helgeson, H. C. (2006) Biomolecules in hydrothermal systems: Calculation of the standard molal thermodynamic properties of nucleic-acid bases, nucleosides, and nucleotides at elevated temperatures and pressures. *Geochim. Cosmochim. Acta* 70:4680–4724.

LaRowe, D. E., and Dick, J. M. (2012) Calculation of the standard molal thermodynamic properties of crystalline peptides. *Geochim. Cosmochim. Acta* 80:70–91.

Larralde, R., Robertson, M. P., and Miller, S. L. (1995). Rates of decomposition of ribose and other sugars: Implications for chemical evolution. *Proc. Natl. Acad. Sci. USA* 92:8158–8160.

Lathe, R. (2004) Fast tidal cycling and the origin of life. *Icarus* 168:18–22.

Lambert, J. F. (2008) Absorption and polymerization of amino acids on mineral surfaces: A review. *Orig. Life. Evol. Biosph.* 38:211–242.

Lemke, K. H., Rosenbauer, R. J., and Bird, D. K. (2009) Peptide synthesis in early Earth hydrothermal systems. *Astrobiology* 9:141–146.

Levy, M., and Miller, S. L. (1998) The stability of the RNA bases: Implications for the origin of life. *Proc. Natl. Acad. Sci. USA* 95:7933–7938.

Li, J., and Brill, T. B. (2003) Spectroscopy of hydrothermal reactions, Part 26: Kinetics of decarboxylation of aliphatic amino acids and comparison with the rates of racemization. *Int. J. Chem. Kinet.* 35:602–610.

Martins, Z., Price, M. C., Goldman, N., Sephton, M. A., and Burchell, M. J. (2013) Shock synthesis of amino acids from impacting cometary and icy planet surface analogues. *Nature Geosci.* 6:1045–1049.

McCord, T. B., Hansen, G. B., Combe, J.-P., and Hayne, P. (2010) Hydrated minerals on Europa's surface: An improved look from the Galileo NIMS investigation. *Icarus* 209:639.

McGrath, M. A., Hansen, C. J., and Hendrix, A. R. (2009) Observations of Europa's tenuous atmosphere. In *Europa*, edited by R. T. Pappalardo, W. B. McKinnon and K. Khurana, The University of Arizona Press, Tucson, AZ, pp. 485–506.

McKinnon, W. B., and Kirk, R. L. (2007) Triton, In *Encyclopaedia of the Solar System (2nd ed.)*, edited by L.-A. McFadden, P. R. Weissman and T. V. Johnson, Academic Press Amsterdam; Boston. pp. 483–502,

Meinert, C., Filippi, J.-J., de Marcellus, P., d'Hendecourt, L. S. and Meierhenrich, U. J. (2012) N-(2-Aminoethyl) glycine and Amino Acids from Interstellar Ice Analogues. *Chem. Plus. Chem.* 77:186–191.

Miyakawa, S., Cleaves, H. J., and Miller, S. L. (2002) The cold origin of life: B. Implications based on pyrimidines and purines produced from frozen ammonium cyanide solutions. *Orig. Life. Evol. Biosph.* 32:209–218.

Moore, J. M., Black, G., Buratti, B., Phillips, C. B., Spencer, J., and Sullivan R. (2009) Surface properties, regolith, and landscape degradation. In *Europa*, edited by R. T. Pappalardo, W. B. McKinnon and K. Khurana, The University of Arizona Press, Tucson, AZ, pp. 329–349.

Muños Carlo, G. M., Meierhenrich, U. J., Schutte, W. A., Barbier, B., Arcones Segovia, A., Rosenbauer, H., Thiemann, W. H.-P., Brack, A., and Greenberg, J. M.

(2012) Amino acids from ultraviolet irradiation of interstellar ice analogues. *Nature* 416:403–406.

Nakashima, S., Brack, A., Maurel, M. C., Maruyama, S., Isozaki, Y., and Windley, B. F. (2001) Introduction to: Geochemistry and the origin of life. In *Geochemistry and the origin of life*, edited by S. Nakashima, S. Maruyama, A. Brack and B. F. Windley, Universal Academy Press, Tokyo, pp.1–13.

Pappalardo, R. T., Belton, M. J. S., Breneman, H. H., Carr, M. H., Chapman, C. R., Collins, G. C., Denk, T., Fagents, S., Geissler, P. E., Giese, B., Greeley, R., Greenberg, R., Head, J. W., Helfenstein, P., Hoppa, G., Kadel, S. D., Klaaen, K. P., Klemaszewski, J. E., Magee, K., McEwen, A. S, Moore, J. M., Moore, W. B., Neukum, G., Phillips, C. B., Prockter, L. M., Schubert, G., Senske, D. A., Sullivan, R. J., Tufts, B. R., Turtle, E. P., Wagner, R., and Williams, K. K. (1999) Does Europa have a subsurface ocean? Evaluation of the geological evidence. *J. Geophys. Res.* 104:24015–24055.

Paranicas. C., Cooper, J. F., Garrett, H. B., Johnson, R. E., and Sturner, S. J. (2009) Europa's radiation environment and its effects on the surface. In *Europa*, edited by R. T. Pappalardo, W. B. McKinnon and K. Khurana, The University of Arizona Press, Tucson, AZ, pp. 529–544.

Pierazzo, E., and Chyba, C. F. (2002) Cometary Delivery of Biogenic Elements to Europa. *Icarus* 157:120-127.

Pizzarello, S., Cooper, G. W., and Flynn, G. J. (2006) The nature and distribution of the organic material in carbonaceous chondrites and interplanetary dust particles. In *Meteorites and the early solar system II*, edited by D. S. Lauretta and H. Y. McSween, The University of Arizona Press, Tucson, AZ, pp. 625-651

Pizzarello, S., and Shock, E. (2010) The organic composition of carbonaceous meteorites: The evolutionary story ahead of biochemistry. Cold Spring Harb. Perspect. Biol. 2:a002105.

Quirico, E., Orthous-Daunay, F. R., Beck, P., Bonal, L., Brunetto, R., Dartois, E., Pino, T., Montagnac, G., Rouzaud, J. N., Engrand, C., and Duprat, J. (2014) Origin of insoluble organic matter in type 1 and 2 chondrites: New clues, new questions. *Geochim. Cosmochim. Acta* 136:80-99.

Rode, B. M. (1999) Peptides and the origin of life. *Peptides* 20:773–786.

Ruiz-Mirazo, K., Pereto, J., and Moreno, A. (2004) A universal definition of life: Autonomy and open-ended evolution. *Orig. Life. Evol. Biosph.* 34:323–346.

Ruiz-Mirazo, K., Briones, C. and Escosura, A. (2014) Prebiotic systems chemistry: New perspectives of the origins of life. *Chem. Rev.* 114:285–366.

Sakata, K., Kitadai, N., and Yokoyama, T. (2010) Effects of pH and temperature on dimerization rate of glycine: Evaluation of favorable environmental conditions for chemical evolution of life. *Geochim. Cosmochim. Acta* 74:6841–6851.

Schmidt, B., Blankenship, D. D., Patterson, G. W., and Schenk, P. M. (2011) Active formation of 'chaos terrain' over shallow subsurface water on Europa, *Nature* 479:502–505.

Sephton, M. A. (2002) Organic compounds in carbonaceous meteorites. *Natural Product Reports* 19:292–311.

Shock, E. L. (1992) Stability of peptides in high-temperature aqueous solutions. *Geochim. Cosmochim. Acta* 56:3481–3491.

Shock, E. L., Oelkers, E. H., Johnson, J. W., Sverjensky, D. A., and Helgeson, H. C. (1992) Calculation of the thermodynamic properties of aqueous species at high pressures and temperatures. *J. Chem. Soc. Faraday Trans.* 88:803–826.

Showman, A. P., Stevenson, D. J., and Malhotra, R. (1997) Coupled orbital and thermal evolution of Ganymede. *Icarus* 129:367–383.

Sotin, C., Grasset, O., and Beauchesne, S. (1998) Thermodynamic properties of high pressure ices: Implications for the dynamics and internal structure of large icy satellites. In: *Solar System Ices*, edited by B. Schmitt, C. de Bergh and M. Festou, Kluwer Academic Press, Dordrecht, pp. 79–96.

Spencer, J., Tamppari, L. K., Martin, T. Z., and Travis, L. D. (1999) Temperatures on Europa from Galileo Photopolarimeter-Radiometer: Nighttime Thermal Anomalies. *Science* 284:1514–1516.

Spencer, J., Pearl, J. C., Segura, M., Flasar, F. M., Mamoutkine, A., Romani, P., Buratti B. J., Hendrix, A. R., Spilker, L. J., and Lopes, R. M. C. (2006) Cassini Encounters Enceladus: Background and the Discovery of a South Polar Hot Spot. *Science* 311:1401–1405.

Stofan, E. R., Elachi, C., Lunine, J. I., Lorenz, R. D., Stiles, B., Mitchell, K. L., Ostro, S., Soderblom, L., Wood, C., Zebker, H., Wall, S., Janssen, M., Kirk, R., Lopes, R., Paganelli, F., Radebaugh, J., Wye, L., Anderson, Y., Allison, M., Boehmer, R., Callahan, P., Encrenaz, P., Flamini, E., Francescetti, G., Gim, Y., Hamilton, G., Hensley, S., Johnson, W. T. K., Kelleher, K., Muhleman, D., Paillou, P., Picardi, G., Posa, F., Roth, L., Seu, R., Shaffer, S., Vetrella, S., and West, R. (2007) The lakes on Titan. *Nature* 445:61–64.

Suda, K., Ueno, Y. Yoshizaki, M., Nakamura, H., Kurokawa, K., Nishiyama, E., Yoshino, K., Hongoh,Y., Kawachi, K., Omori, S., Yamada, K., Yoshida, N., and Maruyama S. (2014) Origin of methane in serpentine-hosted hydrothermal systems: The $CH_4$-$H_2$-$H_2O$ hydrogen isotope systematics of the Hakuba Happo hot spring. *Earth Pla. Sci. Lett.* 386:112–125.

Titus, T., Anderson, J., Capria, M. T., Tosi, F., Prettyman, T., De Sanctis, M. C., Palomba, E., Grassi, D., Capaccioni, F., Ammannito, E., Combe, J.-P., McCord, T. B., Li, J.-Y., Russell, C. T., Raymond, C. A., Toplis, M., and Sykes, M. V. (2012) Comparison of observed surface temperatures of 4 Vesta to the KRC thermal model and possible implications for GRaND observations [abstract 800]. In *7th European Planetary Science Congress,* the IFEMA-Feria de Madrid, Madrid.

Tokano, T., and Neubauer, F. M. (2002) Tidal winds on Titan caused by Saturn. *Icarus* 158:499–515.

Turtle, E. P., Perry, J. E., McEwen, A. S., DelGenio, A. D., Barbara, J., West, R. A., Dawson, D. D., and Porco, C. C. (2009) Cassini imaging of Titan's high-latitude lakes, clouds, and south-polar surface changes. *Geophys. Res. Lett.* 36:L02204.

Vance, S., Harnmeijer, J., Kimura, J., Hussmann, H., DeMartin, B., and Brown, J. M. (2007) Hydrothermal systems in small ocean planets. *Astrobiology* 7:987–1005.

Zahnle, K., Dones, L., and Levison, H. F. (1998) Cratering rates on the Galilean satellites. *Icarus* 136:202–222.

**TABLE CAPTIONS**

**Table 1**

Standard molal thermodynamic data (25 °C and 1 bar) and the revised Helgeson–Kirkham–Flowers equation states for aqueous species.

**Table 2**

Standard molal thermodynamic data (25 °C and 1 bar) and $C_p^o$ power function coefficients for crystalline compounds.

**Table 3**

Physical properties of water ice Ih.

# TABLES

## Table 1

| Species | $\Delta_f G^o$* | $\Delta_f H^o$* | $S^o_{Pr,\,Tr}$† | $C_p^o$† | $V^o$‡ | $a_1$ | $a_2$ | $a_3$¶ | $a_4$ | $c_1$† | $c_2$ | $\omega$ |
|---|---|---|---|---|---|---|---|---|---|---|---|---|
| | | | | | | $\times 10$§ | $\times 10^{-2}$ | | $\times 10^{-4}$# | | $\times 10^{-4}$# | $\times 10^{-5}$ |
| Gly☆ | −88.62 | −122.83 | 37.89 | 9.3 | 43.2 | 11.30 | 0.71 | 3.99 | −3.04 | 28.5 | −8.40 | 0.23 |
| Ala** | −88.81 | −132.50 | 38.83 | 33.6 | 60.4 | 14.90 | 1.74 | 7.16 | −3.69 | 49.5 | −7.00 | 0.18 |
| Leu** | −84.20 | −153.60 | 50.41 | 95.2 | 107.8 | 24.68 | 7.51 | 19.93 | −8.37 | 102.7 | −3.30 | 0.09 |
| GlyGly☆ | −117.00 | −175.69 | 52.75 | 28.0 | 77.4 | 17.66 | 6.71 | 50.07 | −15.91 | 69.3 | −17.71 | 0.59 |
| AlaGly†† | −116.73 | −186.11 | 50.70 | 60.3 | 95.2 | 14.65 | 21.16 | 12.08 | −3.65 | 54.7 | 0.80 | −0.43 |
| LeuGly†† | −110.62 | −202.66 | 72.60 | 118.8 | 145.3 | 21.40 | 34.65 | 13.18 | −4.21 | 98.6 | 6.48 | −0.76 |
| GlyGlyGly☆ | −144.74 | −226.97 | 70.73 | 44.2 | 112.1 | 24.14 | 6.97 | 33.69 | −10.69 | 63.9 | −16.60 | −1.54 |
| Adenine‡‡ | 74.77 | 31.24 | 53.41 | 56.2 | 89.6 | 21.50 | 8.50 | −2.66 | −5.36 | 87.9 | −15.87 | 0.07 |
| Ribose‡‡ | −179.74 | −247.13 | 59.53 | 66.5 | 95.7 | 22.65 | 7.29 | −5.39 | −3.41 | 134.7 | −32.82 | 0.17 |
| Adenosine‡‡ | −46.50 | −148.49 | 87.19 | 120.3 | 170.7 | 39.55 | 12.90 | 8.97 | −8.82 | 163.2 | −20.10 | 0.23 |

*kcal $mol^{-1}$, †cal $mol^{-1}$ $K^{-1}$, ‡$cm^3$ $mol^{-1}$, §cal $mol^{-1}$ $bar^{-1}$, ‖ cal $mol^{-1}$, ¶cal K $mol^{-1}$ $bar^{-1}$, #cal K $mol^{-1}$, ☆Kitadai (2014), ** Dick et al. (2006), ††Shock (1992), ‡‡LaRowe and Helgeson (2006).

## <u>Table 2</u>

| Species | $\Delta_f G^{o*}$ | $\Delta_f H^{o*}$ | $S^o_{Pr,\,Tr}$† | $C_p^o$† | $V^o$‡ | $a$† | $b$§ ×10³ | $c$ ×10⁻⁵ |
|---|---|---|---|---|---|---|---|---|
| Gly | −88.08¶ | −126.22# | 24.74☆ | 23.7☆ | 44.6☆ | 3.56☆ | 67.6☆ | 0☆ |
| Ala | −88.44** | −134.50** | 30.88** | 29.2** | 63.4☆ | 5.77** | 78.6** | 0** |
| Leu☆ | −85.25 | −154.59 | 50.62 | 46.2 | 115.1 | 6.70 | 132.4 | 0 |
| GlyGly☆ | −116.90 | −178.51 | 43.09 | 39.1 | 87.1 | 8.47 | 102.9 | 0 |
| AlaGly☆ | −116.64 | −185.64 | 50.91 | 43.6 | 101.8 | 10.23 | 111.9 | 0 |
| LeuGly☆ | −112.36 | −205.69 | 67.16 | 61.2 | 153.6 | 10.26 | 171.0 | 0 |
| GlyGlyGly☆ | −145.51 | −230.81 | 60.67 | 54.8 | 120.5 | 10.50 | 148.5 | 0 |
| Adenine†† | 71.86 | 23.16 | 36.09 | 34.2 | 91.1 | 0.37 | 113.4 | 0 |
| Ribose†† | −177.66 | −250.29 | 41.99 | 44.7 | 94.4 | −7.40 | 168.8 | 1.58 |
| Adenosine†† | −48.85 | −156.20 | 69.22 | 69.4 | 173.5 | −3.33 | 239.6 | 1.11 |

*kcal mol$^{-1}$, †cal mol$^{-1}$ K$^{-1}$, ‡cm$^3$ mol$^{-1}$, §cal mol$^{-1}$ K$^{-2}$, ‖ cal K mol$^{-1}$, ¶ calculated from $\Delta_f H^o$ and $S^o_{Pr,\,Tr}$ in the table together with values of $S^o_{Pr,\,Tr}$ of the elements from Cox et al., (1989) (see Kitadai (2014)), # Diaz et al. (1992), ☆LaRowe and Dick (2012), ** Helgeson et al. (1998), †LaRowe and Helgeson (2006).

**Table 3**

| | Symbol | Unit | Value |
|---|---|---|---|
| Density* | $\rho$ | kg m$^{-3}$ | 920 |
| Specific heat* | $C_p$ | J K$^{-1}$ kg$^{-1}$ | 7.037T+185.0 |
| Thermal conductivity* | $k$ | W m$^{-2}$ | 488.12/T+0.4685 |
| Thermal expansion coefficient* | $a$ | K$^{-1}$ | $3.0\times(2.5\times10^{-7}\text{T}-1.25\times10^{-5})$ |
| Surface gravity acceleration | $g$ | m s$^{-2}$ | 1.32 |
| Latent heat* | $L$ | kJ kg$^{-1}$ | 284 |
| Melting temperature at P=0† | $T_{m0}$ | K | 271.15 |
| Slope of melting temperature† | $dT_m/dP$ | 10$^{-7}$ K Pa$^{-1}$ | -1.063 |

* Hobbs (1974)

† Sotin et al. (1998)

## FIGURE CAPTIONS

### Figure 1

Gibbs energies of polymerization of 2Gly to GlyGly, Gly and GlyGly to GlyGlyGly, Gly and Ala to AlaGly, and Gly and Leu to LeuGly as a function of temperature for pressure of 1 Pa (upper) and $10^8$ Pa (lower).

### Figure 2

Contour of the Gibbs energy of polymerization of 2Gly to GlyGly as a function of pressure and temperature (solid lines) and expected temperature profiles of Europa's ice crust with 40 km thickness (dashed lines) and 100 km thickness (dotted lines). Surface radius of the moon is 1,565 km. Ice viscosity at the melting point is assumed to be $10^{15}$ Pa s (thick lines) and $10^{14}$ Pa s (thin lines). Hydrostatic pressure corresponding to a depth from the surface of Europa is shown on the right axis (assuming density of 920 kg/m$^3$ and gravity of 1.31 m/s$^2$.

### Figure 3

Gibbs energies of reaction of a nucleoside (adenosine) from a nucleobase (adenine) and sugar (ribose) to adenosine as a function of temperature for a pressure of 1 Pa (solid) and $10^8$ Pa (dashed).

**Author Disclosure Statement.**
No competing financial interests exist.